\documentclass[prl,aps,twocolumn,10pt,showpacs,groupedaddress,superscriptaddress,floatfix,notitlepage]{revtex4-2}
\usepackage{amsmath,amsfonts,amssymb,graphics,graphicx,epsfig,color,times, braket}
\usepackage[utf8x]{inputenc}
\usepackage{color}
\usepackage{bbm, dsfont}
\usepackage{subfigure}
\usepackage{hyperref} 
\usepackage{mathrsfs}
\usepackage{verbatim}
\begin{document}

\title{Suppression of Quantum Correlations in a Clean-Disordered Atom-Nanophotonic Interface}

\author{I Gusti Ngurah Yudi Handayana}
\email{ngurahyudi@unram.ac.id}
\affiliation{Molecular Science and Technology Program, Taiwan International Graduate Program, Academia Sinica, Taiwan}
\affiliation{Department of Physics, National Central University, Taoyuan City 320317, Taiwan}
\affiliation{Institute of Atomic and Molecular Sciences, Academia Sinica, Taipei 10617, Taiwan}

\author{Yi-Lin Tsao}
\affiliation{Institute of Atomic and Molecular Sciences, Academia Sinica, Taipei 10617, Taiwan}
\affiliation{School of Physics, College of Sciences, Georgia Institute of Technology, Atlanta 30332, USA}

\author{H. H. Jen}
\email{sappyjen@gmail.com}
\affiliation{Institute of Atomic and Molecular Sciences, Academia Sinica, Taipei 10617, Taiwan}
\affiliation{Molecular Science and Technology Program, Taiwan International Graduate Program, Academia Sinica, Taiwan}
\affiliation{Physics Division, National Center for Theoretical Sciences, Taipei 10617, Taiwan}

\date{\today}
\renewcommand{\r}{\mathbf{r}}
\newcommand{\f}{\mathbf{f}}
\renewcommand{\k}{\mathbf{k}}
\def\p{\mathbf{p}}
\def\q{\mathbf{q}}
\def\bea{\begin{eqnarray}}
\def\eea{\end{eqnarray}}
\def\ba{\begin{array}}
\def\ea{\end{array}}
\def\bdm{\begin{displaymath}}
\def\edm{\end{displaymath}}
\def\red{\color{red}}
\pacs{}
\begin{abstract}
	Quantum correlations are essential to the emergent behaviors of quantum systems, supporting key phenomena such as localization or delocalization of particles, quantum avalanches in many-body localized systems, and quantum information transfer. In open atom-nanophotonic systems characterized by long-range spin-exchange interactions, we examine the influence of clean system size on high-order quantum correlations among a clean-disordered atomic array with multiple atomic excitations. By initializing the system far from equilibrium, we observe a suppression of quantum correlations for localized atomic excitations in the disordered zone as the clean system size increases, showcasing the delocalization behavior in the high-order spin-exchange processes. The calculation of the entanglement entropy at the interface further substantiates this thermalizing effect. Our results manifest distinct quantum correlations enabled by long-range interactions mediated by the waveguide, enhance the theoretical comprehension of clean-disordered systems, and provide insights to nonequilibrium quantum dynamics in an atom-nanophotonic platform.
\end{abstract}
\maketitle
{\it Introduction}--Quantum correlations are fundamental to emergent behaviors of quantum systems, influencing phenomena such as quantum criticality \cite{Osterloh2002, Mishra2018} and quantum Kibble–Zurek mechanism \cite{Polkovnikov2005, Zurek2005, Keesling2019}. In particular, quantum correlation can act as a probe of the avalanche mechanism  \cite{Leonard2023}, relevant to the fate of many-body localized phase \cite{Bardarson2012, Agarwal2015, Schreiber2015, Vosk2015, Nandkishore2015, Choi2016, Roushan2017, Abanin2019}, which associates with an accelerated transport of thermalized penetration into the localized system \cite{Luitz2017, Thiery2018, Morningstar2022, Sels2022}. This design of an interface with two separated zones of quench dynamics has lead to rich phenomena, in addition to quantum avalanches in a one-dimensional (1D) Bose-Hubbard system \cite{Leonard2023}, involving the observation of Hilbert space fragmentation\cite{Lan2018, Khemani2020, Sala2020, Moudgalya2022, Khudorozhkov2022, Moudgalya2022_2} that defies thermalization in a tilted two-dimensional (2D) Bose-Hubbard model \cite{Adler2024} and the Rayleigh-Taylor instability that results in a turbulent mixture in a binary quantum fluid \cite{Geng2024}. 

Recently, an interface in 1D atomic chains has been proposed theoretically to separate two dissimilar arrays as an alternative platform to study excitation localization to delocalization transitions under single excitation \cite{Wu2024}, resembling the Anderson localization behavior \cite{Anderson1958}. This atom-nanophotonic platform \cite{Mitsch2014, Lodahl2017, Chang2018, Kim2021, Sheremet2023} forms one of the strongly-coupled open quantum systems \cite{Arcari2014, Tiecke2014, Yala2014, Samutpraphoot2020, Dordevic2021} for nonclassical states generations \cite{Tudela2013, Ramos2014, Pichler2015, Jen2021_bound}, revealing photon-photon \cite{Mahmoodian2018, Jeannic2021} or atom-atom correlations \cite{Jen2022_correlation}, and promises scalable quantum computation using graph states \cite{Chien2024}. With individually controlled atoms in optical tweezers, a scalable atomic array coupled to a waveguide can be envisioned to tackle the effect of thermal inclusions on the localized atomic excitations under strong disorders at long time \cite{Zhong2020, Mirza2017, Jen2020_disorder, Fayard2021}. 

This letter focuses on multi-excitation dynamics far from equilibrium in a clean-disordered atom-nanophotonic system with inherent nonlocal spin-exchange couplings \cite{Kien2005,Solano2017}. We probe the impact of variable clean system sizes on the excitation dynamics and obtain the associated quantum correlations using Kubo cumulant expansions \cite{Kubo1962}. By configuring the system with initial states such as the half-Dicke state \cite{Dicke1954} as localized excitations in a disordered region, we numerically calculate the high-order quantum correlations among the clean and disordered zones at sufficiently long time. These correlations are designated for atomic excitations quenched in the clean and disordered sites or distributed in respective zones, where we find significant suppression of these long-time and long-range quantum correlations as the clean zone size increases. This suppression indicates a transition towards excitation delocalization, which is also manifest in a greater entanglement entropy at the interface. Our results elucidate the pivotal influence of the clean system size as thermal inclusions to the excitation localization and enhance the comprehension of the mechanism in excitation delocalization from collective photon-mediated spin-exchange interactions in open atom-nanophotonic platforms. 

\begin{figure}[t]
	\centering
	\includegraphics[width=0.49\textwidth]{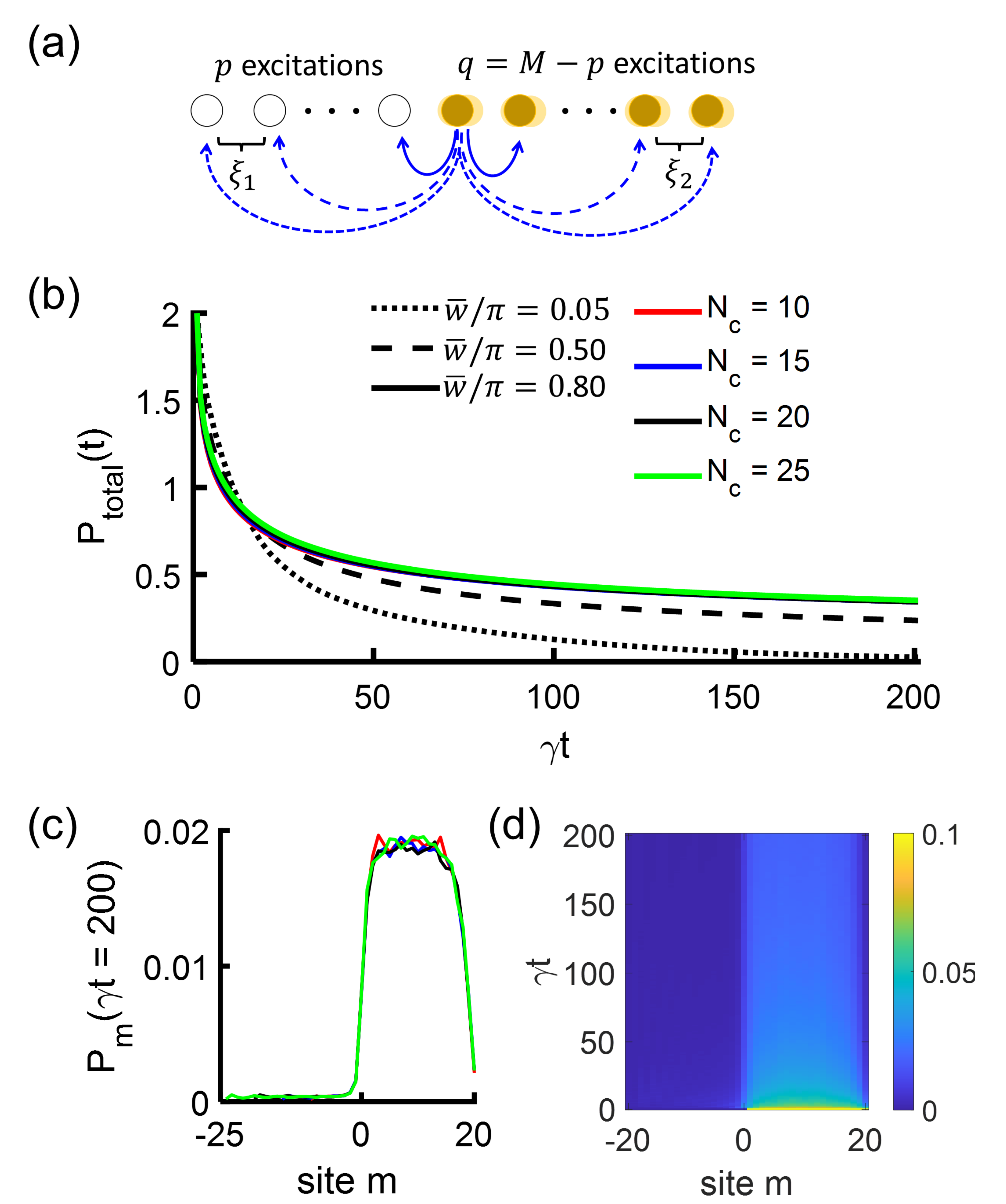}
	\caption{(a) Schematic of an atomic array featuring a clean-disordered interface, with nonreciprocal couplings ($\gamma_L$ and $\gamma_R$). For $M$ excitations, we initialize $p$ excitations in the clean zone and $q$ excitations in the disorder zone, with dimensionless interatomic spacings $\xi_{1,2}$$=$$k_s d_{1,2}$, respectively, at a resonant wavevector $k_s$ on the waveguide with distances $d_{1,2}$. All-to-all couplings are delineated by lines with arrows from the perspective of the atom in the middle. Here $p=0$ as an example with $q$ excitations shared among all sites in the disordered zone. (b) Time evolution of the total population $P_{\rm tot}(t)$ for different clean zone sizes $N_c = 10, 15, 20, 25$ in solid lines under a disorder strength $\bar{w}/\pi = 0.8$ and for different disorder strengths $\bar{w}/\pi = 0.05, 0.5, 0.8$ (dotted, dashed, and solid black) at $N_c = 20$. (c) Cross-sectional population distribution at $\gamma t = 200$ for varying clean zone sizes and (d) population evolution across each site for $N_c=20$ under a disorder strength $\bar{w}/\pi = 0.8$. All plots consider a disorder zone size of $N_d = 20$. All systems in (b), (c) and (d) are initialized with symmetric Dicke state with $M=2$ excitations in the disordered zone ($p=0$, $q=2$) at $\xi_1 = \xi_2 = \pi/4$ and $D=0$.}\label{fig1}
\end{figure}

{\it Theoretical model}--The system under consideration is a 1D atomic array consisting of $N$ atoms and $M$ excitations, divided into two separate regions: a clean zone and a disordered zone, as illustrated in Fig. \ref{fig1}(a). We assign the interface position at the $m = 0$ site, separating two zones, and the dimensionless interparticle distances in the clean and disordered regions are represented by $\xi_1$ and $\xi_2$, respectively. We consider two-level quantum emitters (the ground $|g\rangle$ and the excited state $|e\rangle$ as a spin-$1/2$ system) coupled to a nanophotonic waveguide \cite{Sheremet2023}, where photon-mediated spin-exchange interactions emerge among every pairwise atoms \cite{Pichler2015, Jen2024} due to the guided modes. The system dynamics can be described by a density matrix $\rho$ in Lindblad forms ($\hbar = 1$), with in general left (L)- and right (R)-propagating decay modes \cite{Pichler2015, Lodahl2017}, as
\bea
\frac{d\rho}{dt}=-i[H_{\rm L} +H_{\rm R},\rho] + \mathcal{L}_{\rm L}[\rho] + \mathcal{L}_{\rm R}[\rho].\label{rho}
\eea
The coherent and dissipative terms can be expressed respectively as 
\bea
H_{\rm L(R)}=-i\frac{\gamma_{\rm L(R)}}{2}\sum_{\mu<(>)\nu}^N\sum_{\nu=1}^N (e^{ik_s| r_\mu - r_\nu |}\sigma_\mu^\dagger\sigma_\nu - \rm{H.c.}), \label{H}
\eea
and
\bea
\mathcal{L}_{\rm L(R)}[\rho]=&&-\frac{\gamma_{\rm L(R)}}{2}\sum_{\mu,\nu}^N e^{\mp ik_s(r_\mu - r_\nu)}(\sigma_\mu^\dagger\sigma_\nu\rho + \rho \sigma_\mu^\dagger\sigma_\nu \nonumber\\
&&- 2\sigma_\nu\rho\sigma_\mu^\dagger)\label{L}, 
\eea
where we can define a dimensionless and equal interparticle distance as $\xi=k_s|r_\mu-r_\nu|$. $\sigma_\mu^\dagger\equiv\vert e \rangle_\mu\langle g\vert$ with $\sigma_\mu = (\sigma_\mu^\dagger)^\dagger$ are dipole operators, and $\gamma_{L}\neq \gamma_{R}$ represent the nonreciprocal coupling strengths. The directionality factor can be defined as $D\equiv (\gamma_R - \gamma_L)/\gamma$ \cite{Mitsch2014}, which quantifies the preference of directional spin-exchange interactions under 1D reservoirs \cite{Tudela2013}, with a total decay rate $\gamma = \gamma_R + \gamma_L \equiv 2\vert dq(\omega)/d\omega\vert_{\omega = \omega_{eg}}g^2_{k_s}L$, where $\vert dq(\omega)/d\omega\vert$ indicates an inverse of group velocity with a resonant wavevector $q(\omega)$, the atom-waveguide coupling strength $g_{k_s}$, and the quantization length $L$. $D = \pm 1$ and $0$ present the chiral coupling scheme \cite{Carmichael1993, Gardiner1993, Stannigel2012} and the reciprocal coupling regime, respectively. 

For initial states far from equilibrium, we designate the numbers of $p$ and $q=M-p$ quenched atomic excitations among a total of $M$ excitations in the clean and the disordered zones, respectively. The disorders can be established in atomic position fluctuations, leading to the onsite phase disorders denoted by $W_\mu\in[-\bar w, \bar w]$ on top of $r_\mu$ in Eqs. (\ref{H}) and (\ref{L}), with $\bar w/\pi=[0,1]$. When $p = 0$, all excitations are localized within the disordered zone deep in the localization regime, while the empty sites in the clean zone serve as the thermal baths interacting with the atomic excitations in the disordered zone. When $M$ atoms are quenched to the excited states initially, we need a complete Hilbert space of $C^N_M$ bare states to simulate the system dynamics \cite{Jen2017_MP}, where $C$ denotes the binomial coefficient. Taking $\vert \phi_s \rangle = \sigma^\dagger_j\sigma^\dagger_{k>j} \cdots \sigma^\dagger_l\sigma^\dagger_{m>l}\vert 0\rangle$ as the labeled $s$th bare state basis for $M$ excitations in general, the corresponding probability amplitudes $a_s(t)$ in the state function $|\Psi(t)\rangle$ ($\rho=|\Psi(t)\rangle\langle\Psi(t)|$) can be directly solved from
\bea
\dot{a}_s(t) = \sum_{s=1}^{C^N_M}V_{s,s'}a_{s'}(t)\label{amp},
\eea
where $V_{s,s'}$ represents the interaction matrix between every bare states, encoding photon-mediated dipole-dipole interactions and waveguide-mediated couplings \cite{Jen2022_correlation, Jen2021_bound}. We solve Eq. (\ref{amp}) numerically and calculate the time evolution of every $a_s$. All the results are averaged over $2000$ disorder realizations, which ensures a statistical convergence \cite{SM}.

{\it Excitation transport}--In order to study the excitation localization phenomenon, not all interparticle distances $\xi$ lead to disorder-induced localization. When $\xi=\pi$ or $2\pi$, the atomic excitation dynamics involves populated subradiant and decoherence-free sectors \cite{Jen2020_subradiance}, leading to disorder-induced decays in the excitations instead. Therefore, we focus on small $\xi < \pi/2$ near the superradiant sectors, in a parameter regime for disorder-induced localization under strong disorders. In Figs. \ref{fig1}(b-d), we demonstrate the clean zone size effect on the excitation population dynamics in the disordered zone at long time. We first focus on an initialized half-Dicke state $\vert\Psi(t=0)\rangle$$=$$\sqrt{2}/(\sqrt{N(N-1)})\sum_{i = x_p+1}^{N-1}\sum_{j = i+1}^{N}\sigma_i^\dagger\sigma_j^\dagger\vert 0 \rangle$, where two atomic excitations are equally distributed in the disordered zone ($p=0$). This configuration avoids positional biases of excitations relative to the interface or chain boundaries, and at the time of initialization, the total spin excitations should be conserved, that is $\sum_{m=1}^N P_m(0) = M$, where $P_m\equiv \langle\Psi| \sigma_m^\dag\sigma_m|\Psi\rangle$. In Fig. \ref{fig1}(b), we find that the total populations are largely unaffected by the clean system sizes and remain in the disordered zones when deep in the localization regime. This shows negligible effect of spin-exchange interactions on the excitation transport from the disordered to the clean zones  \cite{Jen2020_disorder, Wu2024} which is further evidenced in the overlapped site-resolved populations at long time under variable clean zone sizes as shown in Fig. \ref{fig1}(c). Moreover, Fig. \ref{fig1}(d) reveals that an absence of population exchanges is established immediately upon initialization, showing the robustness of disorder-induced excitation localization.

This long-time decaying localization of multiple atomic excitations can turn to a faster timescale by a finite $D$, where a larger $D$ directs spin excitations preferentially toward one side of the system \cite{Pichler2015, Lodahl2017}. This asymmetry accelerates the spin transport toward the boundaries, leading to increased excitation losses. Additionally, the presence of unavoidable non-guided modes $\gamma_{\rm ng}$ of the system can further lead to a limited time window ($1/\gamma_{\rm ng}$) for observation of the long-time behaviors \cite{SM}. Therefore, a strong coupling regime $\beta = \gamma / (\gamma + \gamma_{\rm ng})\approx 1$ is required for disorder-induced localization phenomenon and other novel collective radiations \cite{Lodahl2017} or applications like photon routing using atom-nanophotonic interfaces \cite{Sheremet2023}. 

\begin{figure}[b]
	\centering
	\includegraphics[width=0.49\textwidth]{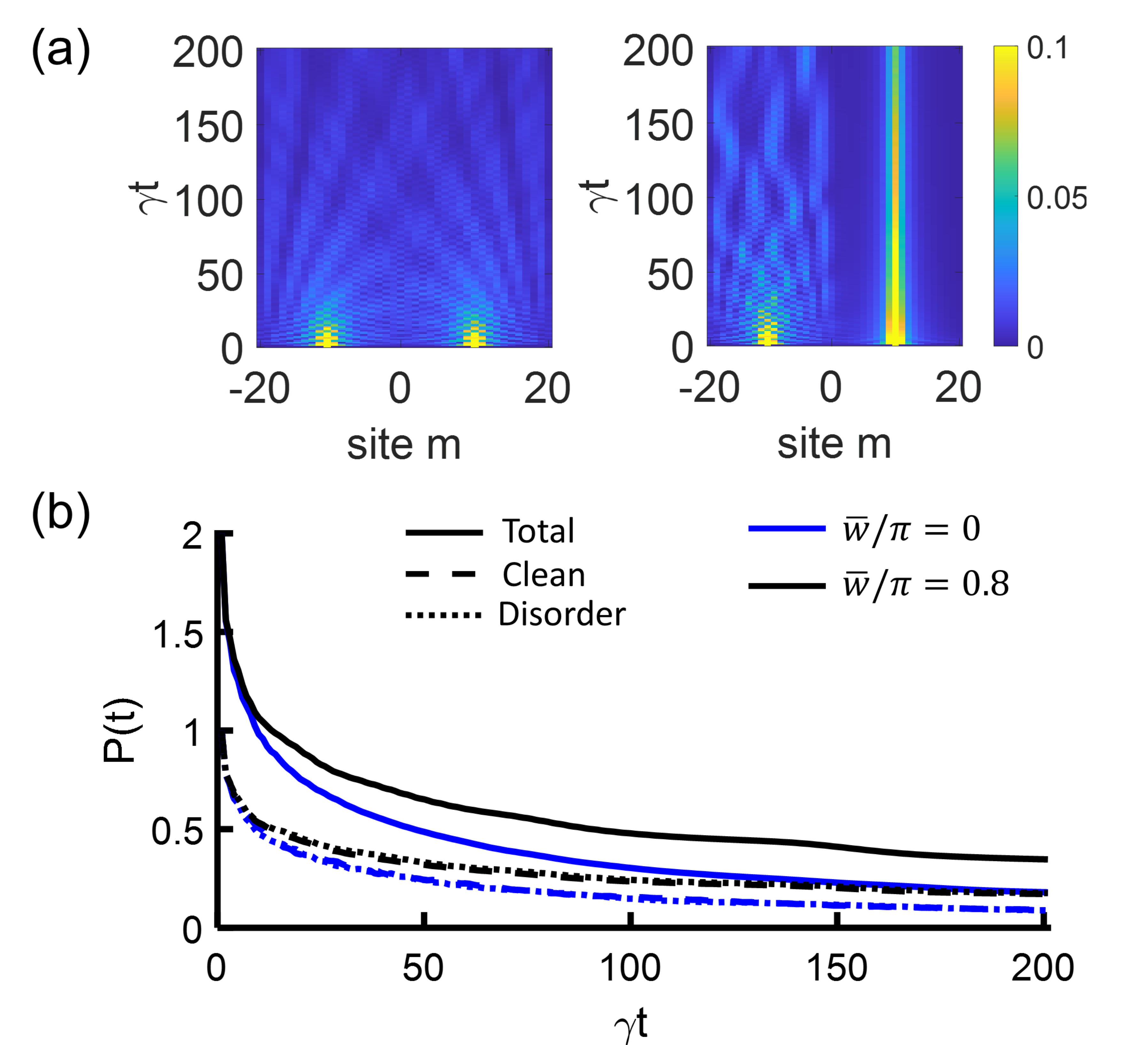}
	\caption{Time evolution of the total excitation populations in an initialization with two atomic excitations with $p=q=1$. (a) Two-dimensional plots of the population dynamics for $\bar{w} = 0$ (left panel) and $\bar{w} = 0.8\pi$ (right panel), illustrating the effects of disorders on population distributions over time. (b) Comparison of the total population (solid lines), populations in the clean (dashed lines) and disordered zones (dotted lines) for $\bar{w} = 0$ (blue) and $\bar{w} = 0.8\pi$ (black). The $N_{c,d}=20$ and the other parameters are the same as in Fig. \ref{fig1}.}\label{fig2}
\end{figure}

To further investigate the population dynamics, we examine the case of $p = 1$ and $q = 1$, where the initial state is prepared with one excitation quenched at position $m = -10$ in the clean zone and the other at $m = 10$ in the disorder zone, as shown in Fig. \ref{fig2}. In the disorder-free case in the left panel of Fig. \ref{fig2}(a), the excitations spread evenly across all sites over time due to the reciprocal coupling, resulting in interference fringes from equal counter-propagating spin excitation transport. In contrast, under a strong disorder in the disordered zone (right panel), the excitation localization emerges, while the excitation in the clean zone spreads freely within its region and does not penetrate the other side. Interestingly, despite the different behaviors of excitation evolution in the clean and disordered zones, the spin-exchange interactions between two zones induce a back-action effect. This effect ensures that the respective total populations in the clean and disordered zones almost overlap over time, regardless of the presence or absence of disorders, as illustrated in Fig. \ref{fig2}(b). This back-action arises because the clean and disordered zones exchange excitation populations, influencing each other and creating balanced dynamics \cite{SM}. Specifically, the localized excitation in the disorder zone, affected by the delocalized excitation in the clean zone, exhibits a reduced population compared to a stand-alone disordered array. Conversely, the excitation in the clean zone, which would typically experience a rapid population decay, retains its population at longer time due to the coupling with the localized excitation in the disordered zone and manifests an effective slowing down in decay at a time longer than the disorder-free case. 

{\it Quantum correlations}--Furthermore, high-order quantum correlations can provide insights and probe the clean-disordered system under multiple excitations. For finite atomic excitations in the clean zone, a larger clean system size promotes entanglement as time evolves and induces intriguing many-body dynamics, which complicates the clean zone size effect on the excitation localization in the disordered zone. Therefore, we focus on empty sites in the clean zone as thermal baths to the disordered zone. 

\begin{figure}[t]
	\centering
	\includegraphics[width=0.48\textwidth]{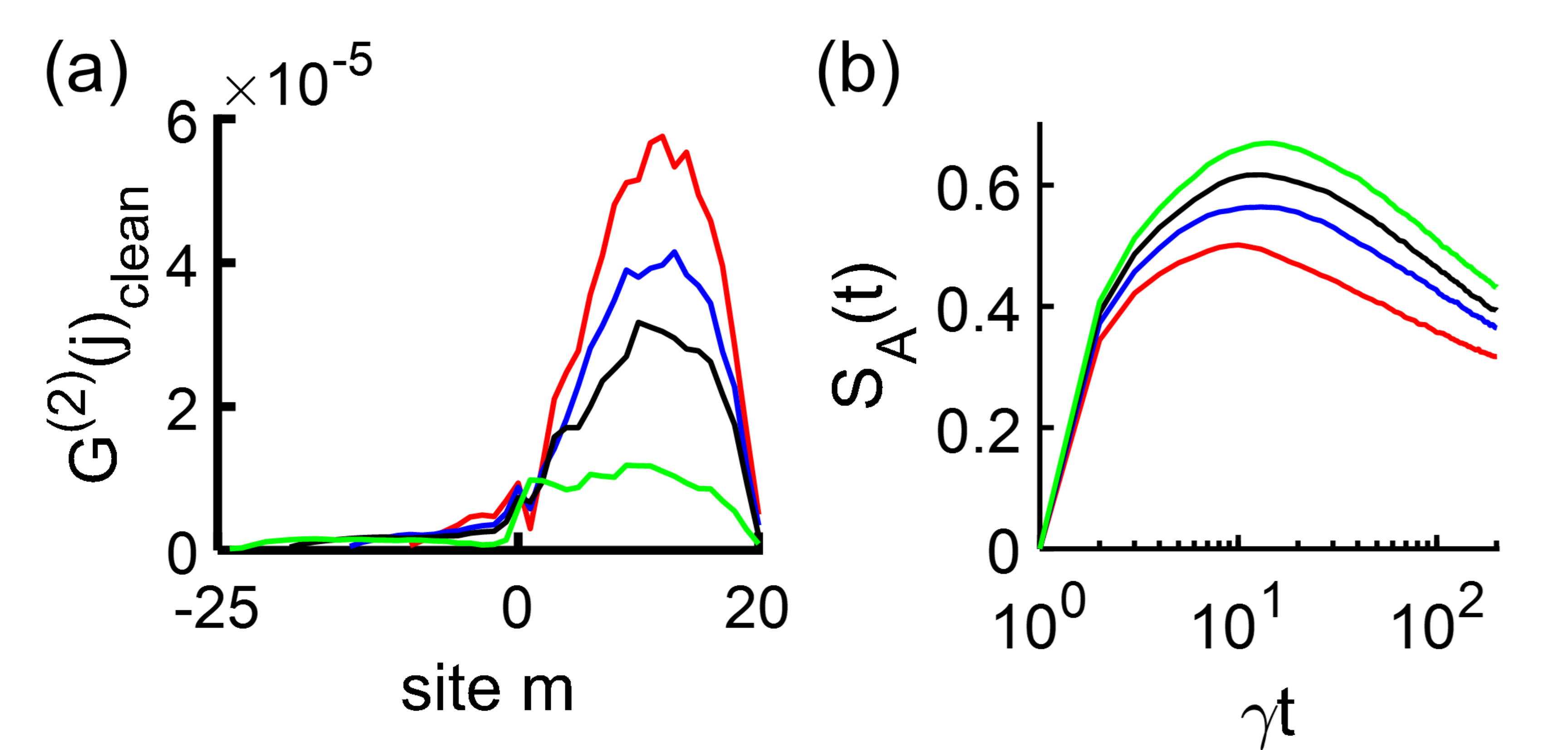}
	\caption{Mean second-order correlations $G^{(2)}(j)_{\rm clean}$ and entanglement entropy $S_A(t)$ for an initialization with $M=2$ excitations, $p=0$, and $D=0$. (a) Suppression of $G^{(2)}(j)_{\rm clean}$ as $N_c$ increases at $\gamma t = 200$. (b) Entanglement entropy $S_A(t)$ with varying clean zone sizes, with a logarithmic scale in time. The line colors for different $N_c$ are the same as in Fig. \ref{fig1}.} \label{fig3}
\end{figure}

Using $M=2$, $N_d=20$, and $p=0$ with a half-Dicke state as the initial excitation, we focus on the strong disorder regime to study the impact of clean system sizes on quantum correlations. The nonclassical correlation between two sites (second-order correlation) in terms of Kubo cumulant expansions \cite{Kubo1962} can be defined as
\bea
G^{(2)}(i,j) = \langle n_i n_j \rangle - \langle n_i \rangle \langle n_j \rangle,
\eea
where $n_j\equiv \sigma_j^\dag\sigma_j$. Specifically, we can compute the mean correlation of the clean zone as \cite{Leonard2023} 
\bea
G^{(2)}(j)_{\rm clean}=\overline{G^{(2)}(i,j)}\vert_{i\in \rm clean},
\eea
which isolates the effect of the clean system on the whole chain. As shown in Fig. \ref{fig3}(a), the mean correlations in the clean zone are extremely low, which reflects limited spin excitations spreading into the clean zone from the disordered zone. By contrast, the mean correlation in the disorder zone is consistently higher than in the clean zone, emphasizing the role of disorders in enhancing localization-induced correlations \cite{SM}. This observation aligns with the pronounced excitation localization in the disorder zone in Fig.\ref{fig1} (c). 

Notably, the mean correlations become suppressed as the clean system size increases, even when the population distributions are kept intact under strong disorders. This feature suggests that the clean zone influences the localized excitations in the high-order processes of thermalization via all-to-all spin-exchange interactions. The facilitated delocalization behavior in quantum correlations can also be compared to a Bose-Hubbard model with nearest-neighbor couplings and local on-site interactions, where thermal inclusions of extra baths penetrate through the interface and thermalize the insulator region with larger penetration depths \cite{Leonard2023}. By contrast, the excitation delocalization here manifests in an overall suppression of quantum correlations with no sign of characteristic length scales of correlations near the interface, due to its initialized distinction between occupied and unoccupied atomic excitations present in the disordered and clean zones, respectively. 

Next, we calculate the entanglement entropy $S_A(t)$ as a complementary measure to quantum correlations to further investigate the delocalization behaviors driven by the clean system size. The entanglement entropy is defined as $S_A(t) \equiv -{\rm Tr}[\rho_A(t)\ln\rho_A]$, where $\rho_A(t) \equiv {\rm Tr}_B[\rho(t)]$ represents the reduced density matrix among the system partitions $A$ and $B$ with a cut at the interface. As shown in Fig. \ref{fig3} (b), we observe an initial rise in the entanglement entropy across all clean system sizes $N_c$ before the effect of disorders kicks in \cite{Wu2024}. However, after $t \sim 2\gamma^{-1}$, deviations of $S_A$ emerge with the entropy growing to its peak value and subsequently decaying due to the excitation dissipation. Notably, as the clean system size increases, the entropy for larger systems consistently remains higher than for smaller ones, while the peak values of $S_A$ delays as $N_c$ increases, showing the timescale for information flow among two zones. This observation implies that larger clean zones allow greater access to distribute spin excitations and to build up quantum correlations as time evolves, enhancing delocalization effect. 

Both the second-order quantum correlations on average $G^{(2)}(j)_{\rm clean}$ and the entanglement entropy at the interface shown in Fig. \ref{fig3} indicate enhanced thermalization of atomic excitations as $N_c$ enlarges. The former characterizes the strength of coincidences of finding the particles within the disordered zone, while the latter identifies the correlations among two separate zones. The suppressed $G^{(2)}(j)_{\rm clean}$ manifests the weakening of localized excitation in the disordered zone, while the increased entropy signifies a trend to thermal equilibrium, supporting the findings of quantum correlations. We note that our results underscore the collective impact of the clean zone on the disordered zone via long-range spin-exchange interactions and highlights the essential role of clean system size in engaging and influencing the overall quantum dynamics, beyond the solely local effects of entropy per particle \cite{SM} seen in systems with short-range interactions \cite{Leonard2023}.

\begin{figure}[b]
	\centering
	\includegraphics[width=0.48\textwidth]{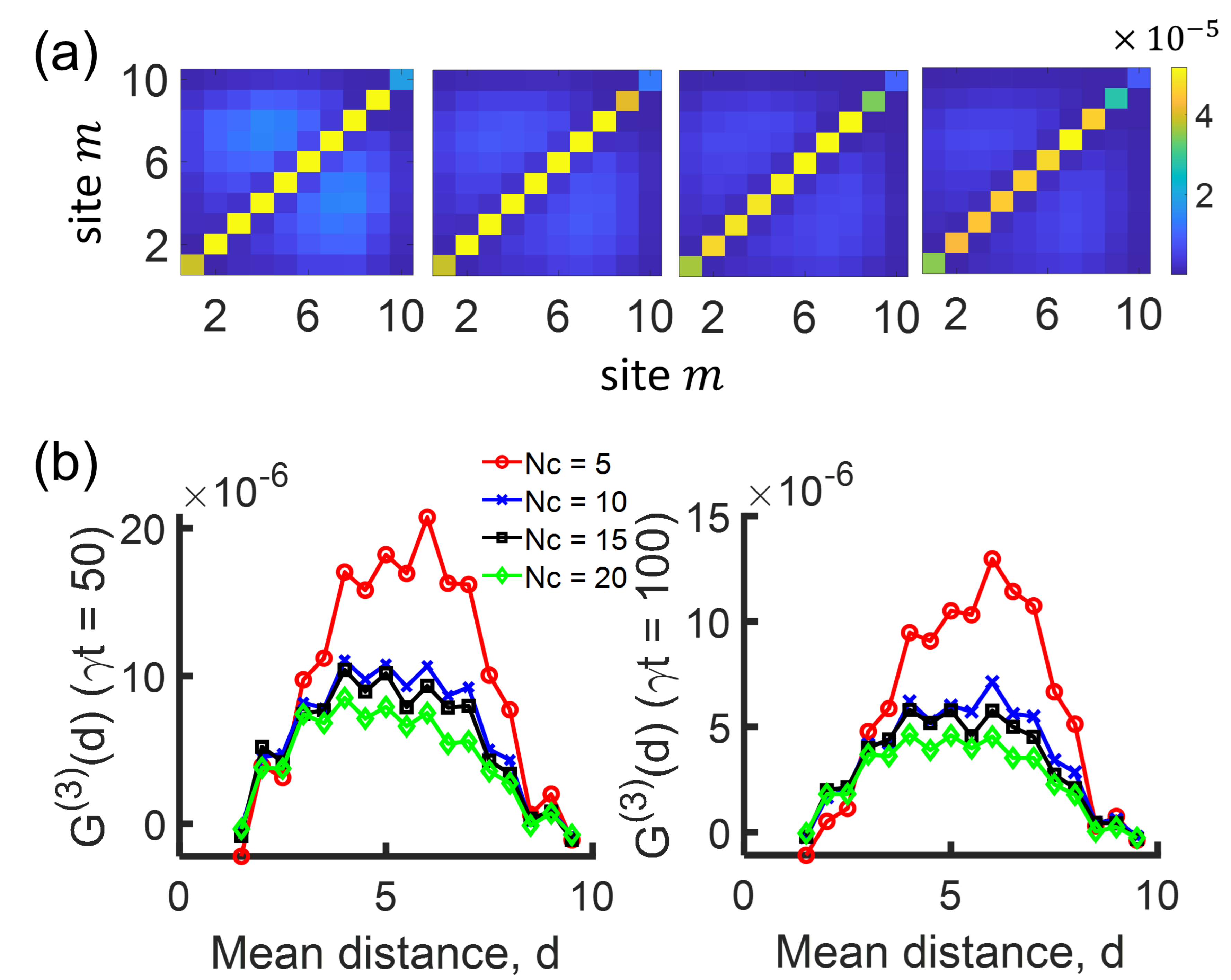}
	\caption{Third-order correlation $G^{(3)}_{(i,j)}$ for an initialization with $M=3$ excitations and $p=0$, among pairs of disordered sites $i$ and $j$, and one clean site $k$ (averaged over all clean sites $k$), in a system with $D=0$, $N_d = 10$, and the disorder strength $\bar{w}/\pi = 0.8$. (a) A 2D heatmap of the average correlation between disordered sites for different clean zone sizes $N_c = 5$, $10$, $15$, $20$ (from left to right) at $\gamma t = 100$. (b) Correlation suppression with increasing clean zone sizes for $N_c = 5$, $10$, $15$, $20$ (red, blue, black and green), averaged over the mean distance from the interface $d = (i+j)/2$, at $\gamma t = 50$ (left panel) and $\gamma t = 100$ (right panel).}\label{fig4}
\end{figure}

Finally, we extend the analysis to multipoint correlations by calculating the third-order correlation function $G^{(3)}(i,j,k)$, using again the Kubo cumulant expansions \cite{Leonard2023, Kubo1962} to quantify nonclassical higher-order quantum correlations. The third-order quantum correlation is defined as 
\bea
G^{(3)}(i,j,k) = &&\langle n_i n_j n_k\rangle -\langle n_i\rangle\langle n_j n_k\rangle-\langle n_j\rangle\langle n_k n_i\rangle\\\notag
&&-\langle n_k\rangle\langle n_i n_j\rangle + 2\langle n_i\rangle\langle n_j \rangle\langle n_k\rangle.
\eea
To analyze the influence of the clean zone, we calculate the mean third-order correlations involving two disordered sites, $i$ and $j$, and one clean site, $k$, averaged over the clean zone, given by
\bea
G^{(3)}(i,j)_{\rm clean} = \overline{G^{(3)}(i,j,k)}\vert_{k\in \rm clean}.
\eea 

In Fig. \ref{fig4}, we focus on the clean system size effect on correlations among two disordered sites from an initialized half-Dicke state with three atomic excitations equally shared in the disordered zone. From the left to the right panels in Fig. \ref{fig4}(a), corresponding to increasing clean system sizes, the mean third-order correlations between two sites in the disordered zone are clearly suppressed, along the diagonal or on the off-diagonal parts. This trend aligns with the findings with the second-order quantum correlations and the entanglement entropy, reinforcing the delocalization effect induced by larger clean system sizes. To further analyze this reduced correlation in the disordered zone, we examine the mean correlation as a function of the mean distance, $d = (i+j)/2$, between these two sites \cite{Leonard2023}. In Fig. \ref{fig4}(b), a plateau of high correlations emerge around $d=5$, corresponding to a balanced distance for two sites equally distributed in the disordered zone of $N_d=10$. Additionally, the third-order correlation exhibits a unique spatial dependence. For two sites near the interface and at the chain boundaries, the correlations become vanishing. This behavior again contrasts with short-range interacting systems, where the mean third-order correlation is strongest near the interface \cite{Leonard2023}. This distinction provides valuable insights into the fundamental differences between systems with nearest-neighbor hopping and long-range spin-exchange couplings, highlighting the unique collective dynamics in atom-waveguide interfaces. As a final remark, we do not observe a speedup in decaying quantum correlations as $N_c$ increases or at longer evolution time. The former can be further clarified in regimes beyond the few atomic excitations but suffers from an exponentially growing dynamical spaces, while the latter is inevitable in generic open quantum systems. 

{\it Experimental feasibility}--Our results are readily simulated and implemented in photonic crystal waveguides coupled with an atomic array \cite{Sheremet2023, Goban2015, Douglas2015} under the strong coupling regime $\beta\sim 1$. These scalable and trapped atoms can be manipulated by optical lattices \cite{Corzo2019} or optical tweezers \cite{Samutpraphoot2020, Dordevic2021}, and the initialized states considered in this work can be created from controlled side excitations \cite{Mitsch2014}. 

{\it Conclusions}--We have provided comprehensive and extensive investigations of the role of clean system sizes in the multi-excitation localization in atom-nanophotonic systems using high-order quantum correlations and entanglement entropy. We uncover a suppression of high-order quantum correlations for localized atomic excitations in the disordered zone as the clean system size increases, facilitating delocalization behaviors in the high-order spin-exchange processes even when the excitation populations are sustained by the strong disorders. This result is also confirmed by the property of entanglement entropy that scales up under larger clean systems. Our results showcase unique quantum correlations which are uniformly suppressed and enabled by long-range spin-exchange interactions, contrasting the behaviors in systems with short-range interactions. These findings underscore the fundamental impact of clean system size and interaction range on quantum dynamics, offering new insights into the interplay between thermal inclusions and the disorder-induced localization. This work not only advances the theoretical understanding of a clean-disordered interface \cite{Handayana2024, Chung2024, Mondal2024}, but also paves the way for designing scalable quantum simulations \cite{Jen2020_steady} that leverage tunable long-range interactions \cite{Xing2022} and structured environments.

We are grateful for helpful discussions with Sumit Goswami, Qian-Rui Huang, and Chun-Chi Wu. We acknowledge support from the National Science and Technology Council (NSTC), Taiwan, under the Grants No. 112-2112-M-001-079-MY3 and No. NSTC-112-2119-M-001-007, and from Academia Sinica under Grant AS-CDA-113-M04. We are also grateful for support from TG 1.2 of NCTS at NTU.  


\begin{widetext}
\section{Supplementary Material for Suppression of Quantum Correlations in a Clean-Disorder Atom-Nanophotonic Interface}

\subsection{Convergence of realization}
\begin{figure}[b]
\centering
\includegraphics[width=0.9\textwidth]{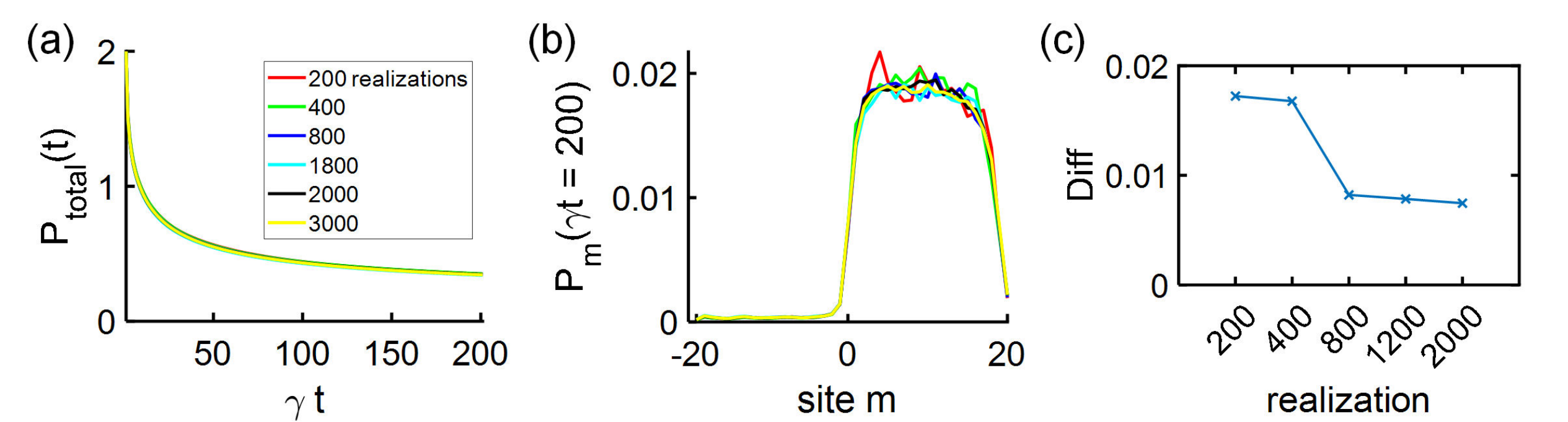}
\caption{Convergence analysis of disorder realizations. (a) Time evolution of the total population for the system using $200$, $400$, $800$, $1200$, $2000$, and $3000$ realizations, represented by solid red, green, blue, cyan, black, and yellow lines, respectively. (b) Site-resolved population at $\gamma t = 200$, illustrating the population distributions across all sites for different realizations. (c) Convergence metric Diff, computed using $3000$ realizations as the reference, showing the deviation across varying realizations. All plots correspond to a system with $N = 40$, $M = 2$, and $p = 0$ (half-Dicke state), with the interface located at site $x_p = 20$.}\label{sup1}
\end{figure}

To ensure convergence of the numerical calculations, in Fig. \ref{sup1} we systematically compare the population dynamics of the system across six different sets of independent disorder realizations: $200$, $400$, $800$, $1200$, $2000$, and $3000$ realizations. The convergence can be quantified by a difference as 
\bea 
{\rm Diff} = \sum_{i=1}^{N} \vert p_i^{(\#1)}(t) - p_i^{(\#2)}(t) \vert\notag, 
\eea 
where $p_i^{(\#1)}(t)$ and $p_i^{(\#2)}(t)$ represent the site populations at time $t$ for two independent disorder realizations. This formula quantifies the deviations between the populations across different realizations to evaluate statistical convergence. For the specific parameters $N = 40$, $M = 2$, and $p = 0$ (corresponding to the half-Dicke state initialization), we observe that the total population is nearly identical across all realization sets. However, only slight variations persist in the site-specific populations due to the inherent randomness of the disorders. Using $3000$ realizations as the reference ($\#2$), by choosing at $\gamma t = 200$, we find that the Diff is reduced and approaches within $2\%$ relative errors for $2000$ realizations, indicating a statistical convergence. This analysis confirms that calculations based on $2000$ realizations provide sufficiently accurate results for the disorder-averaged dynamics.

The calculations presented in this study are constrained by computational resources, particularly memory and processing time, which become increasingly demanding with larger system sizes and excitation numbers. For the case of $M = 3$ excitations, we limit the system size to $N = 30$ atoms to ensure efficient use of memory and computational stability. The simulations are performed on a workstation equipped with an Intel $i9-14900K$ processor and $64$ GB of RAM. Parallel computing is employed, utilizing $7$ cores in MATLAB to optimize the computation time. This parallelization is particularly crucial for handling the large space of basis states, which grows as $C^N_M$. For $M = 2$ excitations, the reduced basis space allows us to extend the calculations to $N = 45$ atoms. This upper limit is chosen to balance the accuracy of the results and the practicality of computational time, particularly for iterative simulations across multiple disorder realizations at long time.

\subsection{Effect of directionality ($D$) and coupling efficiency ($\beta$)}

The primary discussion in the main text assumes perfect coupling to guided modes, as non-guided modes introduce faster decay rates that can obscure localization effects \cite{Sheremet2023}. Figure \ref{sup2} (a) illustrates the effect of non-guided decays, showing limited time window for sustained atomic excitations as the coupling to guided modes decreases (i.e., as $\beta$ becomes smaller). This faster decay prevents clear observation of localization, as excitations dissipate before significant localization dynamics can occur. For this reason, the analysis throughout the main text adopts the assumption of perfect guided mode coupling, $\beta = 1$, to ensure that the effects of localization and delocalization are not masked by rapid decays. In addition to non-guided modes, chirality of couplings can significantly modify the quantum dynamics in atom-nanophotonic systems \cite{Pichler2015, Lodahl2017}. To investigate this effect, systems with finite directionality ($D$) are considered as shown in Fig.~\ref{sup2} (b). A larger $D$ accelerates the decay of excitations toward the end of the atomic chain, primarily due to the boundary effects where directional transport leads to an excitation loss at the system edges. This rapid loss diminishes the time window again to observe long-term dynamics influenced by the clean zone. Consequently, to isolate and study the effects of the clean zone more effectively, the analysis in the main text focuses on reciprocal coupling, $D = 0$, where excitations are exchanged symmetrically among all atoms and boundary-induced losses are minimized. 

\begin{figure}[h]
\centering
\includegraphics[width=0.6\textwidth]{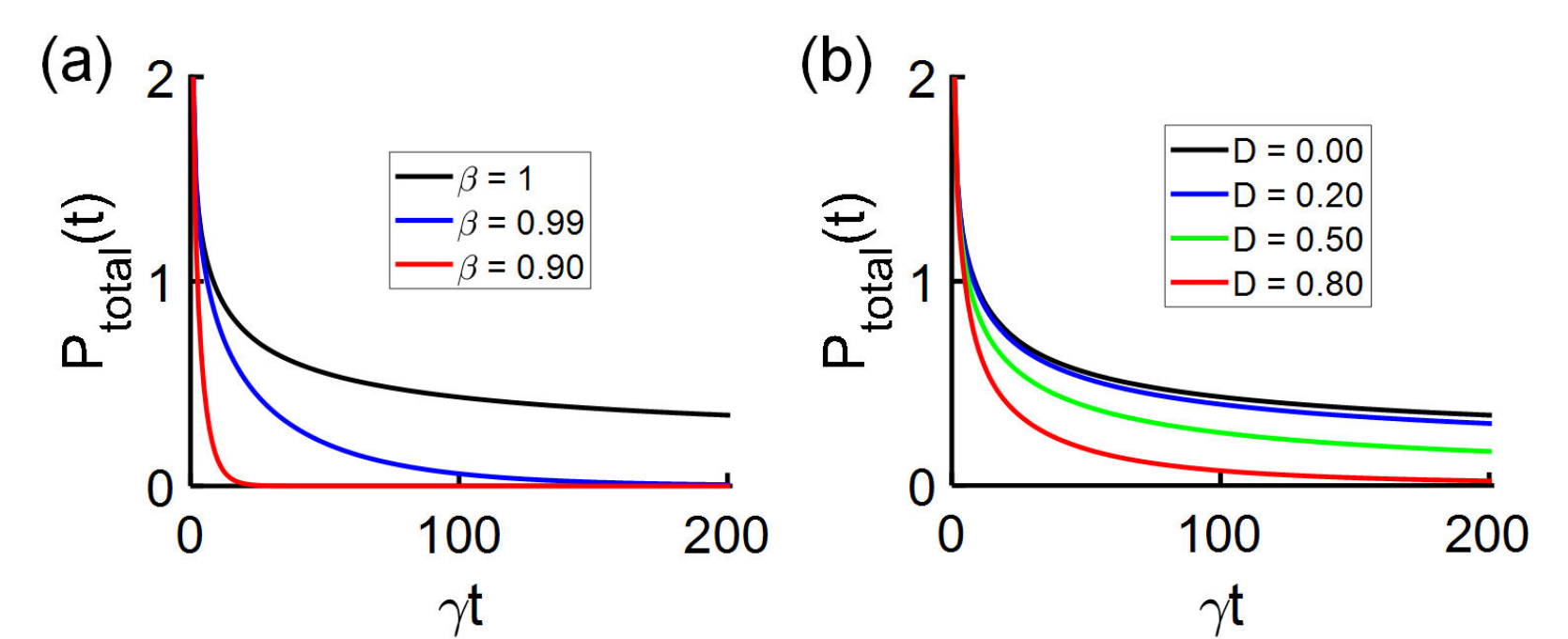}
\caption{Time evolution of the total population under finite non-guided modes and directionality factor $D$. (a) The effect of the non-guided mode in the parameter ($\beta$) on the population dynamics for $\beta = 1$ (solid black), $0.99$ (solid blue), and $0.90$ (solid red). (b) The effect of directionality ($D$) on the population dynamics for $D = 0$ (solid black), $0.2$ (solid blue), $0.5$ (solid green), and $0.8$ (solid red). All plots correspond to a system with $N = 40$, $M = 2$, $N_d = N_c = 20$, $\xi = \pi/4$, and an initial half-Dicke state localized in the disordered zone.}\label{sup2}
\end{figure}

\subsection{Back-action population behavior}

Figure \ref{sup3} demonstrates the back-action effect between clean and disordered zones by comparing three different setups: (a) a system with no disorder, (b) a system with a clean-disorder configuration, and (c) a system with all disorder. All cases are initialized with $p=q=1$ (total atomic excitations $M=p+q=2$) and quenched at sites $m=-10$ and $m=10$, respectively. Numerically, only the basis states $\phi_q$ containing excitations at these sites are initialized. Each case is analyzed by dividing it into comparable regions, such as the clean zone and the disordered zone. With identical parameters for $D$, $\xi$, and $\bar{w}$, the populations of comparable regions across the setups should, in principle, remain the same. For example, the clean zone in case (a), labeled as (I), should have the same population as the clean zone in case (b), labeled as (II). Similarly, the disordered zone in case (b), labeled as (III), should match the disordered zone in case (c), labeled as (IV). However, due to the back-action effect that balances dynamics in clean-disordered arrays, the population dynamics in (b) behaves differently from the results in (a) and (c) as references. The disordered zone (III) enhances the population of the clean zone (II), as evident from a higher population in (II) than in (I). Conversely, the clean zone (II) suppresses the population of the disordered zone in (III), as indicated of a lower population than the one in (IV). Despite these spin-exchange interactions among clean and disordered zones, the populations of (II) and (III) remain almost balanced, showing significant overlapped populations throughout the time evolution. 

\begin{figure}[t]
\centering
\includegraphics[width=0.85\textwidth]{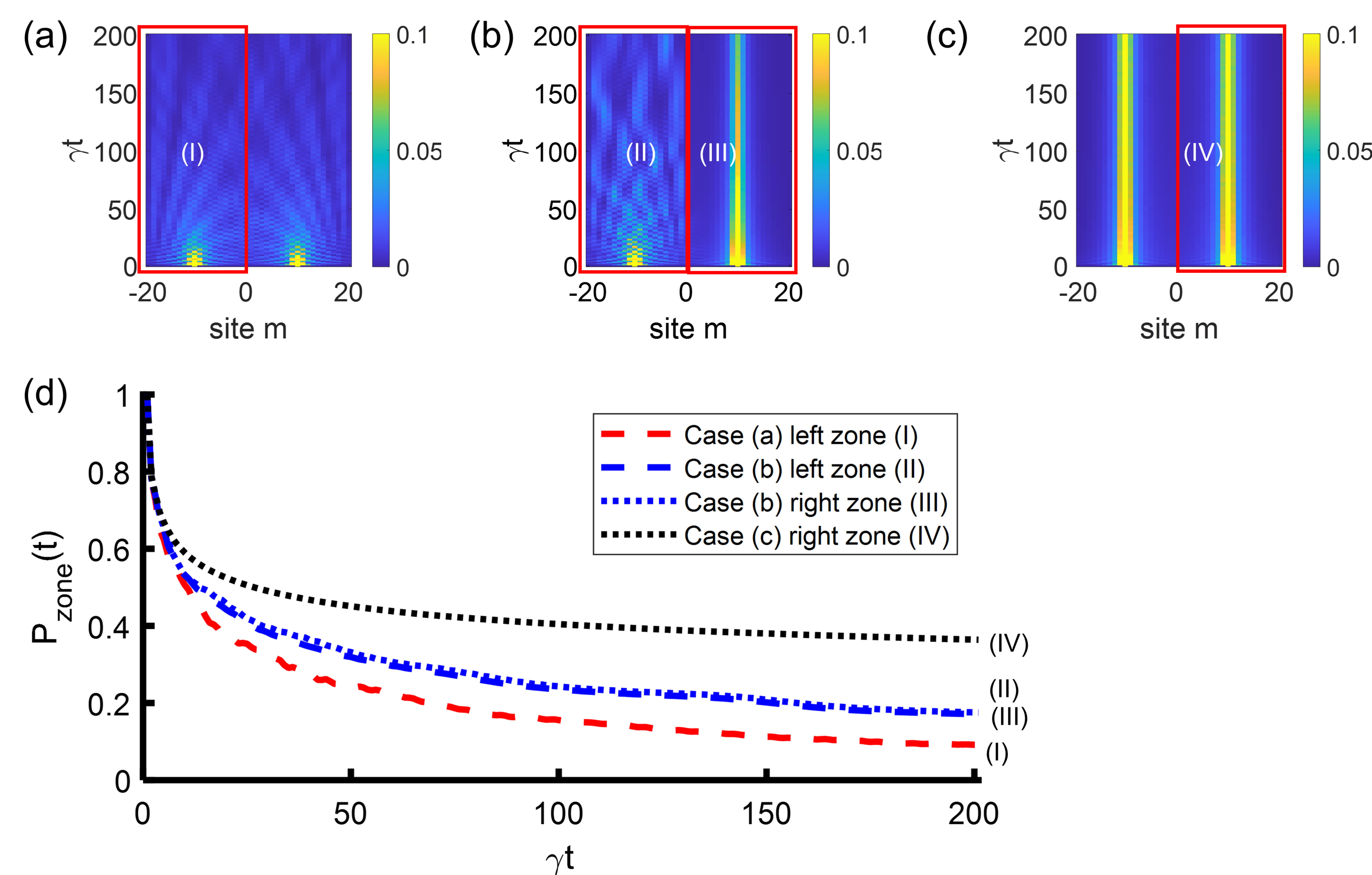}
\caption{Back-action mechanism of population dynamics between the clean zone and the disordered zone. (a–c) Time evolution of site populations under different system configurations: (a) no disorder (fully clean system), (b) mixed clean-disorder system with a clean zone (II) and a disordered zone (III), and (c) fully disordered system. The colormaps show the localized or delocalized nature of excitations over time. (d) Comparison of population dynamics in the clean and disordered zones for different setups: (I) fully clean system, (II) clean zone in the mixed system, (III) disordered zone in the mixed system, and (IV) fully disordered system. In all plots, we consider $N = 40$, $M = 2$, $N_d = N_c = 20$, $\xi = \pi/4$, and initial state quenched with respective single excitations in the site $m=-10$ and $m=10$.}\label{sup3}
\end{figure}


\subsection{Averaging of $G^{(2)}(m,m')$}

To ensure that the observed suppression of quantum correlations is not an artifact of the averaging process over the clean zone, where positive and negative correlation values between sites $i$ and $j$ ($i\in clean zone$) could potentially cancel each other out and result in artificially low average values, we performed additional checks on atom-atom correlations. Figure \ref{sup4} presents 2D plots of the quantum correlations between site $m$ and the other sites $m'$ at $\gamma t = 200$ for varying clean zone sizes: $Nc = 10$, $15$, $20$, and $25$, shown in panels (a), (b), (c), and (d), respectively. The plots clearly demonstrate that within the clean zone ($m, m'< 0$), the correlations are uniformly positive, with no significant negative contributions. This observation confirms that the suppression of correlations as the clean system size increases is an inherent property of the system rather than a computational artifact resulting from the averaging process. Furthermore, the absence of negative correlations within the clean zone highlights the robustness of the results and reinforces the conclusion that increasing the clean zone size reduces quantum correlations in the disordered region due to the interplay between excitation localization in the disordered zone and delocalization dynamics assisted in the clean zone.
 
\begin{figure}[h]
\centering
\includegraphics[width=0.9\textwidth]{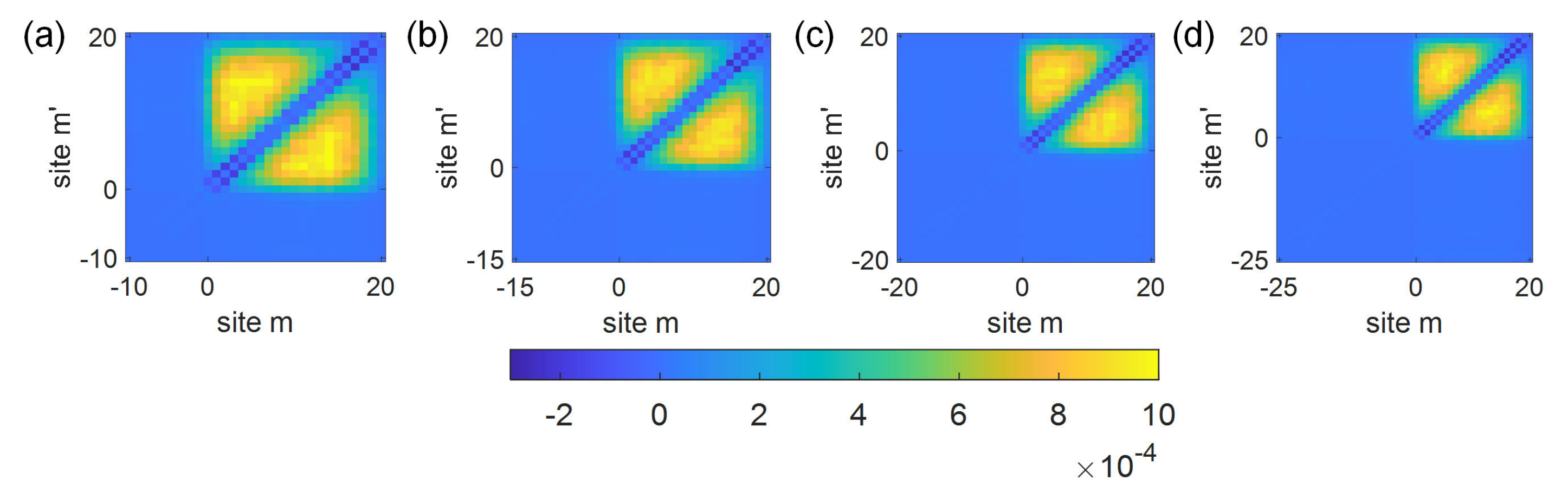}
\caption{Two-dimensional plots of $G^{(2)}(m,m')$ at $\gamma t = 200$ for varying clean zone sizes. The clean zone sizes are (a) $N_c = 10$, (b) $N_c = 15$, (c) $N_c = 20$, and (d) $N_c = 25$, with the disordered zone size fixed at $N_d = 20$. Each plot corresponds to a system with $N = 40$, $M = 2$, and $\xi = \pi/4$}\label{sup4}
\end{figure}

\subsection{Entropy per particle ($S_i(t)$)}

The entropy per particle is defined as $S_i=-\sum_{n_i}p(n_i)\ln p(n_i)/\langle \bar{n_i}\rangle$ where $p(n_i)$ represents the probability distribution of the atomic population at site $i$ \cite{Leonard2023}. For a two-level system, $n_i$ can take values $0$ or $1$, with $p(n_0) = 1 - p(n_1)$. For two excitations $M=2$ as initialized half-Dicke state, this measure of entropy quantifies the degree of uncertainty or fluctuations in the populations at the site $i$, normalized by the average population $\langle \bar{n}_i \rangle$. To explore the spatial and temporal evolution of the entropy, we calculate $S_i(t)$ at three distinct positions: (i) near the interface, (ii) in the bulk of the disordered zone, and (iii) at the chain's end. Figure \ref{sup5} illustrates the time evolution of $S_i(t)$ for these three positions. The results show that different positions, either near the interface, in the bulk of the disordered zone, or close to the boundary, yield varying entropy values, reflecting the heterogeneity in population dynamics across the system. For example, sites in the bulk exhibit lower entropy showing localization, while those on the chain's end or near the interface display higher values, more prone to delocalization effect. Interestingly, as the clean system size increases, there is no significant change in the entropy per particle for any site positions. This behavior highlights a fundamental difference between systems with long-range interactions and those with nearest-neighbor interactions. 
 
\begin{figure}[h]
\centering
\includegraphics[width=0.25\textwidth]{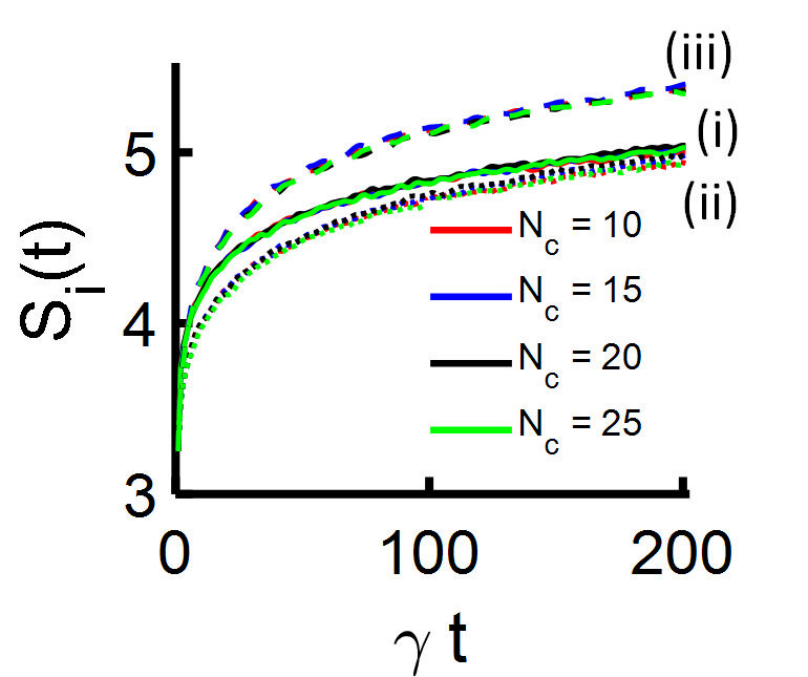}
\caption{Entropy per particle $S_i(t)$ for different positions in the system with varying clean zone sizes under half-Dicke state initialization. The clean zone sizes are $N_c = 10$, $15$, $20$, and $25$, represented by solid red, blue, black, and green lines, respectively. The entropy is calculated and plotted for three specific sites: near the interface ($m = 2$, labeled as (i)), in the bulk region ($m = 10$, labeled as (ii)), and near the end of the chain ($m = 18$, labeled as (iii)).}\label{sup5}
\end{figure}

\end{widetext}
\end{document}